\documentclass[12pt]{article}
\usepackage[english]{babel}
\usepackage[pdftex]{color}
\usepackage{amsmath}
\usepackage{graphicx}
\usepackage{hyperref}
\usepackage{amsmath}
\usepackage{amsfonts}
\usepackage{amssymb}

\begin{document}

\title{On finite $N=1,2$ BRST transformations: Jacobians and Standard Model with gauge-invariant Gribov horizon\thanks{%
Talk presented at  SFP'16, 06 June -- 11 June, 2016, TSU, Tomsk, Russia}}

\author{A.A. Reshetnyak\thanks{Institute of Strength Physics and Materials Science of SB RAS, 634021, Tomsk,
Russia; e-mail: reshet@ispms.tsc.ru}$\ $ and $\ $P.Yu. Moshin\thanks{%
Tomsk State University, Department of Physics, 634050, Tomsk, Russia; e-mail: moshin@phys.tsu.ru}
}
\date{}
\maketitle

\begin{abstract}
We review the concept and properties of finite field-dependent BRST and BRST-antiBRST
transformations introduced in our recent study \cite{Reshetnyak,MR1,MR2,MR3,MR4,MR5,MR6}
for gauge theories.
Exact rules for calculating the Jacobian of a corresponding change of variables in the partition function
are presented. Infrared peculiarities under $R_\xi$-gauges in the Yang--Mills theory and Standard Model
are examined in a gauge-invariant way with an appropriate horizon functional and unaffected local
 $N=1,2$ BRST symmetries.
\end{abstract}

\section{Introduction}

BRST transformations \cite{BRST1,BRST2} for gauge theories in Lagrangian formalism
with a \emph{field-dependent} \emph{(FD)} Grassmann parameter $\mu$ were considered
for the first time within the BV method \cite{BV} in the infinitesimal case, so as to prove
the invariance of the partition function $Z_{\Psi } $: $Z_{\Psi }=Z_{\Psi +\delta \Psi }$
with respect to small variations of the gauge (in terms of the gauge fermion $\Psi $) under
the choice $\mu =-\frac{\imath}{\hbar }\delta \Psi $. FD BRST transformations in the case
of a \emph{finite} functional parameter were introduced in Yang--Mills (YM) theories,
within the family of generalized $R_{\xi }$-gauges \cite{JM}, as a sequence of infinitesimal
FD BRST transformations. Developed initially as a special $N=1$\ SUSY transformation,
and being a change of the field variables
$\phi ^{A}\rightarrow \phi ^{A\prime }=\phi ^{A}+\delta _{\mu }\phi ^{A}$
in the integrand of $Z_\Psi= \int d\phi \exp \Big\{\frac{\imath }{\hbar }S_{\Psi}(\phi )\Big\}$
with a quantum action $S_{\Psi }(\phi )$, BRST transformations were extended by means
of antiBRST transformations \cite{aBRST1,aBRST2} in YM theories to the case of $N=2$ BRST
(BRST-antiBRST) transformations (in YM \cite{aBRST4} and general gauge theories
\cite{BLT1}), parametrized by an $\mathrm{Sp}(2)$ doublet of Grassmann parameters,
$\mu _{a}$, $a=1,2$.

The study of \cite{Sorella}, which suggested to analyze so-called \emph{soft BRST symmetry breaking}
in YM theories, with account taken of the Gribov problem \cite{Gribov} in the infrared region
of field configurations, discussed in the Zwanziger recipe \cite{Zwanziger} by adding
a BRST-non-invariant horizon functional $H$ to the quantum action, attracted the attention of A.A.R. (whose e-mail
communication of 11.03.2011 initiated the joint study \cite{llr1} together with P.M.~Lavrov and O.~Lechtenfeld).
Among the results of \cite{llr1} in the field-antifield formalism related to \cite{Sorella}, an equation was derived
in softly broken BRST symmetry (SB BRST) for a bosonic term $M(\phi ,\phi^{\ast })$ added to the quantum
action $S_{\Psi }(\phi ,\phi ^{\ast })$ of a general gauge theory. The validity of this equation, involving
the generating functional of Green's functions $Z_{\Psi,M}(J,\phi ^{\ast })$ depending on sources $J_{A}$, preserves
the gauge-independence of the respective vacuum functional $Z_{\Psi ,M}(0)$ and the effective action,
depending on external antifields, $\Gamma_{M}=\Gamma _{M}(\phi ,\phi ^{\ast })$, and evaluated
on the extremals,
\begin{eqnarray}
\hspace{-0.7em} &\hspace{-0.7em}&\hspace{-0.7em}\Big[M_{A}(\textstyle\frac{%
\hbar }{\imath }\overrightarrow{\partial }_{(J)},\phi ^{\ast })\Big(\hspace{%
-0.1em}\overrightarrow{\partial }{}^{\ast A}-\frac{\imath }{\hbar }M^{\ast
A}(\frac{\hbar }{\imath }\overrightarrow{\partial }_{J},\phi ^{\ast })%
\hspace{-0.1em}\Big)\delta \Psi ({\textstyle\frac{\hbar }{\imath }}%
\overrightarrow{\partial }_{(J)})+\delta M(\frac{\hbar }{\imath }%
\overrightarrow{\partial }_{(J)},\phi ^{\ast })\hspace{-0.1em}\Big]Z_{\Psi
,M}(J,\phi ^{\ast })\hspace{-0.1em}=\hspace{-0.1em}0  \label{softbreak} \\
\hspace{-0.7em} &\hspace{-0.7em}&\hspace{-0.7em}\Longrightarrow \textstyle Z_{\Psi
,M}(0)=Z_{\Psi +\delta \Psi ,M+\delta M}(0)\quad \mathrm{and}\quad \delta \Gamma _{M}%
\big|_{\Gamma _{M,A}=0}=0, \ \Gamma _{M,A} = \Gamma _{M}\frac{\overleftarrow{\delta }}{\delta \phi ^{A}}%
=\Gamma _{M}\overleftarrow{\partial }_{A},  \label{softbreak1} \\
\hspace{-0.7em} &\hspace{-0.7em}&\hspace{-0.7em} \mathrm{with} \ Z_{\Psi ,M}(J,\phi ^{\ast })\hspace{-0.15em}=\textstyle\hspace{-0.3em}%
\int \hspace{-0.25em}d\phi  \exp \left\{ \hspace{-0.25em}\frac{\imath }{%
\hbar }(S_{\Psi }(\phi ,\phi ^{\ast })\hspace{-0.1em}+\hspace{-0.1em}M(\phi
,\phi ^{\ast })+J_{A}\phi ^{A})\hspace{-0.25em}\right\} \ \mathrm{and} \ \Gamma _{M}=\textstyle\frac{%
\hbar }{\imath }\ln Z-J_{A}\phi ^{A}, \label{genfun}
\end{eqnarray}%
where it is assumed that
$[M_{A},M^{\ast A}]\equiv \lbrack M\overleftarrow{\partial }_{A},\overrightarrow{\partial }^{\ast A}M]$, and $\phi ^{A}
=\frac{\hbar }{\imath }\overrightarrow{\partial }_{(J)}^{A}\ln Z$ are average
fields. (\ref{softbreak}) has the same form when the horizon functional $H(A)$ for YM fields $A^{\mu n}(x)$ is used
as $M(\phi ,\phi ^{\ast })$. In terms of the vacuum expectation value, in the presence of external sources $J_{A}$
and with a given gauge $\Psi $, relation (\ref{softbreak}) acquires the form
\begin{equation}
\left\langle \hspace{-0.1em}%
\delta M+M\overleftarrow{s}\textstyle\frac{\imath }{\hbar }\delta \Psi (\phi
)\hspace{-0.1em}\right\rangle =\left\langle \hspace{-0.1em}\delta M-M%
\overleftarrow{s}\mu (\delta \Psi )\hspace{-0.15em}\right\rangle \hspace{%
-0.1em}=\hspace{-0.1em}0,\ \mathrm{for}\,\overleftarrow{s}\hspace{-0.2em}=%
\hspace{-0.1em}\overleftarrow{\partial }_{A}
\overrightarrow{\partial }{}^{\ast A}S_{\Psi }:\delta _{\mu }\phi ^{A}%
\hspace{-0.1em}\equiv \hspace{-0.1em}\phi ^{A}\overleftarrow{s}\hspace{-0.2em%
}\mu ,  \label{softbreakm}
\end{equation}%
with $\overleftarrow{s}$ being the generator of BRST transformations. In fact, the horizon functional in the family
of $R_{\xi }$-gauges for small $\xi$ was derived explicitly in \cite{llr1} by Eq.~(5.20) therein. By using FD BRST transformations with a small odd-valued parameter, this was established in \cite{Reshetnyak}. A.A.R. drew the attention of his coauthors (communication of 05.12.2011 to P.M.~Lavrov) in \cite{llr1} to the study of \cite{mandalr} which attempted to use FD BRST transformations \cite{JM} for relating the vacuum functionals of YM and GZ (Gribov--Zwanziger) theories in the same gauge. Explicit calculations of the functional Jacobian for a change of variables induced by FD BRST transformations in YM theories with a finite parameter $\mu $ were made in \cite{LL1} so as to establish the gauge-independence of $Z_{\Psi ,M}|_{M=0}$ under a finite change of the gauge.

The present article reviews the study of finite $N=1,2$ BRST transformations (including the case of FD parameters)
and the way they influence the properties of the quantum action and path integral in conventional quantization.
We suggest a quantum action for the YM theory and the Standard Model with an $N=1,2$ BRST-invariant
horizon functional in terms of gauge-invariant transverse fields $(A^{h})^{n}_{\mu}(x)$, with the initial BRST
symmetry under $R_{\xi }$-gauges, in a way different from the recipe of \cite{Pereira}. We use the DeWitt
condensed notation and the conventions of \cite{Reshetnyak,MR1}, e.g., $\epsilon (F)$,
$\overleftarrow{\partial }^{A},\overrightarrow{\partial }_{A}^{\ast }$ and $\overrightarrow{\partial
}_{(J)}^{A}$ are used to denote the respective value of the Grassmann parity of a quantity $F$ and derivatives
with respect to (anti)field variables $\phi ^{A},\phi_{A}^{\ast }$ and sources $J_{A}$. The raising and lowering
of $\mathrm{Sp}\left( 2\right)$ indices, $\big(\overleftarrow{s}^{a},\overleftarrow{s}_{a}\big)=\big(%
\varepsilon ^{ab}\overleftarrow{s}_{b},\varepsilon _{ab}\overleftarrow{s}{}^{b}\big)$, are carried out
by the antisymmetric tensor $\varepsilon^{ab}$, $\varepsilon ^{ac}\varepsilon _{cb}=\delta _{b}^{a}$,
$\varepsilon^{12}=1$.

\section{$N=1,2$ finite BRST transformations}

\label{BRSTproposal} 

Finite FD BRST transformations for the integrand in (\ref{genfun}) at $J=M=0$
\begin{equation}\label{se}
\delta _{\mu }\phi ^{A}%
\hspace{-0.1em}\equiv \hspace{-0.1em}\phi ^{A}\overleftarrow{s}_e, \ \ \overleftarrow{s}_{e}= {\overleftarrow{%
\partial }}_{A}S_{\Psi }^{\ast A} \ \mathrm{with} \ (\overleftarrow{s}_{e})^{2}={\overleftarrow{\partial }}%
_{A}(S_{\Psi }^{\ast A}{\overleftarrow{\partial }}_{B})S_{\Psi }^{\ast
B}\neq 0,
\end{equation}
with a finite Grassmann parameter $\mu (\phi ,\phi^{\ast })$, depending on external antifields $\phi _{A}^{\ast }$,
$\epsilon(\phi _{A}^{\ast })+1=\epsilon (\phi ^{A})=\epsilon _{A}$, and internal fields\footnote{The variables
$\phi ^{A}$ contain the classical fields $A^{i}$, $i=1,..,n$, with gauge transformations
$\delta A^{i}=R_{\alpha }^{i}(A)\xi^{\alpha}$, $\alpha =1,..,m<n$, as well as the ghost, antighost,
and Nakanishi--Lautrup fields $C^{\alpha},\bar{C}^{\alpha},B^{\alpha}$,
$\epsilon (A^{i},\xi ^{\alpha },C^{\alpha },\bar{C}^{\alpha },B^{\alpha })$= $\big({\epsilon }_{i}$,
${\epsilon }_{\alpha }$, ${\epsilon }_{\alpha }+1$, ${\epsilon }_{\alpha }+1$, ${\epsilon }_{\alpha}\big )$,
along with additional towers of fields, depending on the (ir)reducibility of the theory.}
$\phi^{A}$, were introduced in \cite{Reshetnyak} and made it possible to solve the problem
of SB BRST symmetry in general gauge theories. The master equation
for $S_{\Psi }$, $\Delta \exp\left\{ \frac{\imath }{\hbar }S_{\Psi }\right\} =0$ with
$\Delta=(-1)^{\epsilon _{A}}{\overrightarrow{\partial }}_{A}{\overrightarrow{%
\partial }}{}^{\ast A}$, reflects the absence of nilpotency for the generator
$\overleftarrow{s}_{e}$, which reduces at $\phi ^{\ast }=0$ to the usual generator
$\overleftarrow{s}$ of BRST transformations.

Construction of finite $N=2$ BRST Lagrangian transformations solving
the same problem within a suitable quantization scheme (starting from YM
theories) was problematic in view of BRST-antiBRST-non-invariance
for the gauge-fixed quantum action $S_{F}$ in a form more than linear
in $\mu _{a}$, $S_{F}(g_{l}(\mu _{a})\phi )=S_{F}(\phi )+O(\mu _{1}\mu _{2})$,
with the gauge condition encoded by a gauge boson $F(\phi )$.
This problem was solved by finite $N=2$ BRST transformations in
an Abelian supergroup form, $\{g(\mu _{a})\}$, using an appropriate
set of variables $\Gamma ^{p}$, according to \cite{MR1}
\begin{equation}
\hspace{-0.1em}\left\{ G\left( \Gamma g(\mu _{a}) \right) =G\left( \Gamma
\right) \ \mathrm{and}\ G\overleftarrow{s}{}^{a}=0\right\} \hspace{-0.1em}%
\Rightarrow g\left( \hspace{-0.1em}\mu _{a}\hspace{-0.1em}\right) =1+%
\overleftarrow{s}{}^{a}\mu _{a}+\textstyle\frac{1}{4}\overleftarrow{s}{}^{2}\mu ^{2}=\exp \left\{ \hspace{-0.1em}\overleftarrow{s%
}{}^{a}\mu _{a}\hspace{-0.1em}\right\} ,  \label{bab}
\end{equation}%
where $G(\Gamma)$ is an arbitrary regular functional; $\mu ^{2}\equiv \mu _{a}\mu ^{a}$,
$\overleftarrow{s}{}^{2}\equiv \overleftarrow{s}{}^{a}\overleftarrow{s}{}_{a}$, and $\overleftarrow{s}{}^{a}$
are the generators of BRST-antiBRST transformations\footnote{The
transformations $\Gamma ^{p}\to \Gamma^{p}g(\mu _{a}) $, however, cannot be presented in terms of
an $\exp $-like relation for an $\mathrm{Sp}(2)$ doublet of functional parameters $\mu _{a}(\Gamma)$,
due to $\mu _{a}\overleftarrow{s}_{b}\neq 0$.} in the space of $\Gamma ^{p}$.

In YM theories, the construction of finite $N=2$ BRST transformations (\ref{bab}) uses an explicit form
of generators  $\overleftarrow{s}{}^a$  (satisfying
$\{\overleftarrow{s}{}^{a},\overleftarrow{s}{}^{b}\}\hspace{-0.15em}=\hspace{-0.15em}0$)
in the space of fields $\phi ^{A}\hspace{-0.15em}$=$\hspace{-0.05em}(A^{i},\hspace{-0.15em}C^{\alpha },
\hspace{-0.15em}\bar{C}^{\alpha}$, $B^{\alpha })$ arranged in $\mathrm{Sp}(2)$-symmetric tensors,
$(A^{i},C^{\alpha a},B^{\alpha })$ = $\big(A^{\mu m},C^{ma},B^{m}\big)$, as follows \cite{MR1}:
\begin{eqnarray}
\hspace{-0.5em} &\hspace{-0.5em}&S_{F}(\phi )=S_{0}(A)-\textstyle\frac{1}{2}%
F_{\xi }\overleftarrow{s}{}^{2},\ S_{0}(A)=-\frac{1}{4}\int \hspace{-0.2em}%
d^{D}x\,G_{\mu \nu }^{m}G^{m\mu \nu },\ G_{\mu \nu }^{m}=\partial _{\lbrack
\mu }A_{\nu ]}^{m}+f^{mnl}A_{\mu }^{n}A_{\nu }^{l},  \label{variat} \\
\hspace{-0.5em} &\hspace{-0.5em}&\ F_{\xi }(\phi )=\textstyle\frac{1}{2}\int
d^{D}x\ \big(- A_{\mu }^{m}A^{m\mu }+\frac{\xi }{2}\varepsilon
_{ab}C^{ma}C^{mb}\big)\  \Leftrightarrow  R_\xi-\mathrm{gauges} ,  \label{F(A,C)} \\
\hspace{-0.5em} &\hspace{-0.5em}&\Delta A_{\mu }^{m}=D_{\mu }^{mn}C^{na}\mu
_{a}-\textstyle\frac{1}{2}\left( D_{\mu }^{mn}B^{n}+\frac{1}{2}%
f^{mnl}C^{la}D_{\mu }^{nk}C^{kb}\varepsilon _{ba}\right) \mu ^{2}\ ,
\label{DAmm} \\
\hspace{-0.5em} &\hspace{-0.5em}&\Delta B^{m}=-\textstyle\frac{1}{2}\left(
f^{mnl}B^{l}C^{na}+\frac{1}{6}f^{mnl}f^{lrs}C^{sb}C^{ra}C^{nc}\varepsilon
_{cb}\right) \mu _{a}\ ,  \label{DBm} \\
\hspace{-0.5em} &\hspace{-0.5em}&\Delta C^{ma}=\left( \varepsilon ^{ab}B^{m}-%
\textstyle\frac{1}{2}f^{mnl}C^{la}C^{nb}\right) \mu _{b}-\textstyle\frac{1}{2%
}\left( f^{mnl}B^{l}C^{na}+\textstyle\frac{1}{6}%
f^{mnl}f^{lrs}C^{sb}C^{ra}C^{nc}\varepsilon _{cb}\right) \mu ^{2},
\label{DCma}
\end{eqnarray}
for $\eta _{\mu \nu }=\mathrm{diag}(-,+,\ldots ,+)$ and the totally
antisymmetric~$su(\hat{N})$ structure constants $f^{mnl}$, $l,m,n=1,\ldots
,\hat{N}{}^{2}-1$).\footnote{$N=2$ and $N=1$ BRST-invariant
actions of YM theories coincide only in Landau gauge, $\xi =0$.}
In general gauge theories, such as reducible theories or theories
with an open gauge algebra, the corresponding space of triplectic variables
$\Gamma _{{tr}}^{p}=(\phi ^{A},\phi _{Aa}^{\ast },\bar{\phi}_{A},\pi ^{Aa},
\lambda ^{A})$ in the $\mathrm{Sp}(2)$-covariant Lagrangian quantization scheme
\cite{BLT1} contains, in addition to $\phi ^{A}$, also 3 sets of antifields
$\phi _{Aa}^{\ast}$, $\bar{\phi}_{A}$, $\epsilon (\phi _{Aa}^{\ast },\bar{\phi}_{A})
=(\epsilon _{A}+1,\epsilon _{A})$, as sources to BRST, antiBRST and
mixed BRST-antiBRST transformations, and 3 sets of Lagrangian multipliers
$\pi ^{Aa},\lambda ^{A}$, $\epsilon (\pi ^{Aa},\lambda ^{A})=(\epsilon_{A}+1,\epsilon _{A})$,
introducing the gauge. The corresponding generating functional of Green's functions,
$Z_{F}(J)= \int \mathcal{I}_{\Gamma _{{tr}}}^{\left( F\right) }(J)$,
with the bosonic functional $W(\phi, \phi^*_a, \bar{\phi})$ being related
to the gauge-invariant classical action $S_0(A)$ as $W(\phi, 0,0)= S_0 (A)$,
\begin{equation}
\hspace{-0.5em}Z_{F}(J)=\hspace{-0.2em}\int \hspace{-0.1em}d\Gamma
\hspace{-0.1em}
\exp \hspace{-0.1em}\left\{ \hspace{-0.15em}\big (\imath /\hbar \big)%
\Big[W+\phi _{a}^{\ast }\pi ^{a}+\bar{\phi}\lambda -\textstyle\frac{1}{2}F%
\overleftarrow{U}{}^{2}+J\phi \Big]\hspace{-0.15em}\right\} ,
\overleftarrow{U}^{a}={\overleftarrow{\partial }}_{A}\pi ^{Aa}+\varepsilon^{ab}
{\overleftarrow{\partial }_{Ab}^{(\pi )}}\lambda ^{A}, \label{z(0)}
\end{equation}%
is invariant at $J=0$ with respect to finite $N=2$ BRST transformations
(for constant $\mu _{a}$) in the space of $\Gamma _{{tr}}^{p}$, which are
obtained from (\ref{bab}) with a functional $G_{{tr}}=G(\Gamma _{{tr}}^{p})$:
\begin{align}
& \Gamma _{{tr}}^{p}\rightarrow \Gamma _{{tr}}^{\prime
p}=\Gamma _{{tr}}^{p}\left( 1+\overleftarrow{s}{}^{a}\mu _{a}+\textstyle%
\frac{1}{4}\overleftarrow{s}{}^{2}\mu ^{2}\right) \equiv \Gamma _{{tr}%
}^{p}g(\mu _{a})\Longrightarrow \mathcal{I}_{\Gamma _{{tr}}g(\mu
_{a})}^{\left( F\right) }(0) =\mathcal{I}_{\Gamma _{{tr}}}^{\left( F\right) }(0),
\label{Gamma_fin} \\
& \hspace{-0.5em}\mathrm{where}\ \overleftarrow{s}{}^{a}\hspace{-0.15em}=%
\hspace{-0.1em}\Big(\hspace{-0.15em}{\overleftarrow{\partial }}_{A},{%
\overleftarrow{\partial }}{}_{(\phi ^{\ast })}^{Aa},{\overleftarrow{\partial
}}_{(\bar{\phi})}^{A},{\overleftarrow{\partial }}_{Ab}^{(\pi )}\hspace{-0.1em%
}\Big)\hspace{-0.1em}\Big(\hspace{-0.1em}\pi ^{Aa},W_{,A}(-1)^{\epsilon
_{A}},\varepsilon ^{ab}\phi _{Ab}^{\ast }(-1)^{\epsilon _{A}+1},\varepsilon
^{ab}\lambda ^{A}\hspace{-0.1em}\Big)^{T}\hspace{-0.4em},\{\overleftarrow{s}{%
}^{a},\overleftarrow{s}{}^{b}\}\hspace{-0.1em}\neq \hspace{-0.1em}0,  \notag
\\
& \ \mathrm{provided\ that}\ \left( \Delta ^{a}+(\imath
/\hbar )\varepsilon ^{ab}\phi _{Ab}^{\ast }{\overrightarrow{\partial }}_{(%
\bar{\phi})}^{A}\right) \exp \left\{ \frac{\imath }{\hbar }W\right\} =0,\
\mathrm{for}\ \Delta ^{a}=(-1)^{\epsilon _{A}}{\overrightarrow{\partial }}%
_{A}{\overrightarrow{\partial }}{}^{\ast Aa}.  \label{eqme}
\end{align}%
%
%
%
%
%
%
%
%
%
%
%
%
%
%
%
%
%
%
%
%
%
%
%
%
%
%
%
%
%
%
%
%
\section{Jacobians of FD $N=1, 2$ BRST transformations}
\label{Jacobian}

The Jacobian induced by a change of variables $\phi ^{A}\rightarrow \phi ^{\prime A}
=\phi ^{A}(1+\overleftarrow{s}_{e}\mu )$ is given by \cite{Reshetnyak}
\begin{eqnarray}
\hspace{-0.7em} &\hspace{-0.7em}&\hspace{-0.5em}\mathrm{Sdet}\left\Vert \phi
^{\prime }{}^{A}\overleftarrow{\partial }_{B}\right\Vert =\exp \hspace{-0.1em%
}\left\{ \hspace{-0.1em}\mathrm{Str}\,\mathrm{ln}\hspace{-0.1em}\left( \hspace{%
-0.1em}\delta _{B}^{A}+(S_{\Psi }^{\ast A}\mu )\overleftarrow{\partial }_{B}%
\hspace{-0.1em}\right) \hspace{-0.1em}\right\} =\exp \hspace{-0.1em}\bigg\{%
\hspace{-0.1em}\mathrm{Str}\hspace{-0.15em}\sum_{n=1}\hspace{-0.2em}\frac{%
(-1)^{n+1}}{n}\hspace{-0.2em}\bigg((S_{\Psi }^{\ast A}\mu )\overleftarrow{%
\partial }_{B}\hspace{-0.2em}\bigg)^{n}\hspace{-0.1em}\bigg\}  \nonumber \\
\hspace{-0.7em} &\hspace{-0.7em}&\hspace{-0.5em}=\ \big(1+\mu \overleftarrow{%
s}_{e}\big)^{-1}\Big\{1+\overleftarrow{s}_{e}\mu \Bigr\}\Big\{1+\bigl(\Delta
S_{\Psi }\bigr)\mu \Bigr\} = J_{\mu(\phi)}   \label{jacobianres}
\end{eqnarray}%
and reduces, in a rank-$1$ theory with a closed gauge algebra, $[\Delta
S_{\Psi },\overleftarrow{s}{}^{2}]=[0,0]$, where$\ \overleftarrow{s}_{e}=%
\overleftarrow{s}$, to the form $\mathrm{Sdet}\left\Vert \Phi ^{\prime
}{}^{A}\overleftarrow{\partial }_{B}\right\Vert \ =\ \big(1+\mu
\overleftarrow{s}\big)^{-1}$, which is the same as in YM theories \cite{LL1}. The
Jacobian (\ref{jacobianres}) allows one to solve the problem of SB BRST
symmetry in general gauge theories \cite{Reshetnyak} and was examined in
detail \cite{MR4} for an equivalent representation of $Z_{\Psi ,M}(J,\phi^{\ast })$
with BRST transformations $\Gamma ^{p}\rightarrow \Gamma^{p\prime }
=\Gamma ^{p}(1+\overleftarrow{{s}}\mu )$, for $\mu (\Gamma )$ and
$\Gamma ^{p}\overleftarrow{{s}}=(\phi ^{A},\tilde{\phi}_{A}^{\ast },\lambda
^{A})\overleftarrow{{s}}=(\lambda ^{A},S\overleftarrow{\partial }_{A},0)$,
in an extended space $\Gamma ^{p}$ of fields $\phi ^{A}$, internal
antifields $\tilde{\phi}_{A}^{\ast }$, and Lagrangian multipliers $\lambda
^{A}$ for Abelian hypergauge conditions, $G_{A}(\phi ,\phi ^{\ast })=\phi
_{A}^{\ast }-\Psi (\phi )\overleftarrow{\partial }_{A}$, with the result
\begin{eqnarray}
\hspace{-0.5em} &\hspace{-0.5em}&Z_{\psi ,M}(J,\phi ^{\ast })= \textstyle\int d{\Gamma }%
\exp \left\{ \frac{\imath }{\hbar }S({\phi },\tilde{\phi}{}^{\ast })+%
\mathcal{G}_{A}\big({\phi },\tilde{\phi}^{\ast \prime }+\phi ^{\ast }\big)%
\lambda ^{A}+M({\phi },{\phi }{}^{\ast })+J\phi \right\} ,  \label{genfunbv}\\
\hspace{-0.5em} &\hspace{-0.5em}&
J^{BV}_{\mu(\Gamma)} = \mathrm{Sdet}\left\Vert \Gamma ^{p\prime }\overleftarrow{\partial }%
_{q}^{\Gamma }\right\Vert =\big(1+\mu \overleftarrow{{s}}\big)^{-1}\Big\{1+\overleftarrow{s}\mu \Bigr\}\Big\{1+%
\bigl(\Delta S\bigr)\mu \Bigr\}.
\label{fdbrst}
\end{eqnarray}%
The Jacobian (\ref{fdbrst}) coincides with (\ref{jacobianres}), except for the $U$-exact term $\overleftarrow{U} = \overleftarrow{s}\big|_{\phi,\lambda}$, $\overleftarrow{U}^2=0$, with the hypergauge
$G_{A}= (\phi^*_A\phi^A+\Psi)\overleftarrow{U}$, which is $\Delta$-exact, thus making the Jacobian
$J_{\mu(\phi)}$ unique.

We suggest a so-called \emph{soft nilpotency condition}, $\tilde{\mu}(\overleftarrow{s}_e)^2 =0$,
for the parameter $\tilde{\mu}(\phi)$ which transforms the set of $G_{w}=\{g(\mu(\phi)):
\mu(\overleftarrow{s}_e)^2 =0\}$ into a group with the  Jacobian
\begin{equation}\label{jacobweak}
J_{\tilde{\mu}(\phi)} = J_{\mu(\phi)}\big|_{G_{w}} =\big(1+\mu \overleftarrow{%
s}_{e}\big)^{-1}\Big\{1+\bigl(\Delta S_{\Psi }\bigr)\mu \Bigr\},
\end{equation}
being more general than in a rank-$1$ theory and formally identical with $J^{BV}_{\mu(\phi,\lambda)}$; see
(\ref {fdbrst}).

For $N=2$ BRST transformations in YM theories, the technique of
calculating the Jacobian was first examined in the case of functionally-dependent
parameters $\mu _{a}=\Lambda (\phi )\overleftarrow{s}_{a}$ with an
even-valued functional $\Lambda $  in \cite{MR1}. The
result is given by, $\phi^{\prime }{}^{A}\equiv \phi^{A}g(\Lambda
(\phi )\overleftarrow{s}_{a})$,
\begin{eqnarray}
&&\ J_{\Lambda (\phi )\overleftarrow{s}_{a}}=\mathrm{Sdet}\left\Vert \phi
^{\prime }{}^{A}\overleftarrow{\partial }_{B}\right\Vert =\exp \hspace{-0.1em%
}\left\{ \hspace{-0.1em}\mathrm{Str}\,\mathrm{ln}\left( \delta
_{B}^{A}+M_{B}^{A}\right) \right\} ,\ \ \mathrm{for}\ \
M_{B}^{A}=P_{B}^{A}+Q_{B}^{A}+R_{B}^{A}  \label{jacobianN2YM} \\
&&\ =\phi ^{A}\overleftarrow{s}^{a}(\mu _{a}\overleftarrow{\partial }%
_{B})+\mu _{a}\big[(\phi ^{A}\overleftarrow{s}^{a})\overleftarrow{\partial }%
_{B}-\textstyle\frac{1}{2}(\phi ^{A}\overleftarrow{s}^{2})(\mu ^{a}%
\overleftarrow{\partial }_{B})\big](-1)^{\epsilon _{A}+1}+\textstyle\frac{1}{%
4}\mu ^{2}(\phi ^{A}\overleftarrow{s}^{2}\overleftarrow{\partial }_{B}),
\notag \\
&&\ \mathrm{Str}(P+Q+R)^{n}=\mathrm{Str}(P+Q)^{n}+C_{n}^{1}\mathrm{Str}%
P^{n-1}R,\ \ \mathrm{for}\ \ C_{n}^{k}=n!/k!(n-k)!,
\label{jacobianN2YMrules} \\
&&\ \mathrm{Str}(P+Q)^{n}=\left\{
\begin{array}{l}
\mathrm{Str}P^{n}+n\mathrm{Str}P^{n-1}Q+C_{n}^{2}\mathrm{Str}P^{n-2}Q^{2},\
n=2,3, \\
\mathrm{Str}P^{n}+n\textstyle\sum_{k=0}^{2}\mathrm{Str}P^{n-k}Q^{k}+K_{n}%
\mathrm{Str}P^{n-3}QPQ,\ n>3%
\end{array}%
\right.   \label{P+Q} \\
&&\ \Longrightarrow J_{\Lambda (\phi )\overleftarrow{s}_{a}}=\exp \hspace{%
-0.1em}\Big\{\hspace{-0.1em}\,\sum_{n=1}(-1)^{n-1}n^{-1}\mathrm{Str}%
(P_{B}^{A})^{n}\Big\}=\left( 1-\textstyle\frac{1}{2}\Lambda \overleftarrow{s%
}{}^{2}\right) ^{-2},  \label{jacobianN2YMres}
\end{eqnarray}%
where $K_{n}=\left[ \frac{n+1}{2}-2\right] C_{n}^{1}+\big((n+1)\,\mathrm{mod}%
\,2\big)C_{\left[ \frac{n}{2}\right] }^{1}$, with $[x]$ being the integer
part of $x\in \mathbb{R}$. For functionally-independent FD parameters, $\mu
_{a}(\phi )\neq \Lambda \overleftarrow{s}{}_{a}$, the above algorithm (\ref%
{jacobianN2YM})--(\ref{jacobianN2YMres}) involves a generalization of (\ref%
{P+Q}), examined separately for odd- and even-valued $n$, which leads
to \cite{MR5}
\begin{equation}
\hspace{-0.5em}\hspace{-0.5em}J_{\mu _{a}}=\exp \hspace{-0.1em}\Big\{\hspace{%
-0.1em}\,\mathrm{tr}\sum_{n=1}(-1)^{n-1}n^{-1}\mathrm{Str}(P_{B}^{A})^{n}%
\Big\}=\exp \hspace{-0.1em}\left\{ \hspace{-0.1em}\,-\mathrm{tr}\ln
(e+m)\right\} ,\ m_{b}^{a}=\mu _{b}\overleftarrow{s}{}^{a},
\label{jacobianN2YMgen}
\end{equation}%
where $\left( e\right) _{b}^{a}$ and $\mathrm{tr}$ denote $\delta _{b}^{a}$
and trace over $\mathrm{Sp}(2)$ indices. The Jacobian (\ref{jacobianN2YMgen})
is generally not BRST-antiBRST-exact; however, it is identical at $\mu
_{a}=\Lambda \overleftarrow{s}_{a}$ with  $J_{\Lambda
\overleftarrow{s}_{a}}$ (\ref{jacobianN2YMres}),
due to $\mathrm{tr}\,m_{b}^{a}=-\Lambda
\overleftarrow{s}{}^{2}$.
In general gauge theories (\ref{z(0)})--(\ref{eqme}), the calculation of
Jacobians induced by FD $N=2 $ BRST transformations was first carried out
in \cite{MR2,MR4} with functionally-dependent parameters $\mu _{a}=\Lambda
(\phi ,\pi ,\lambda )\overleftarrow{U}_{a}$, the restricted generators $%
\overleftarrow{U}{}^{a}=\left. \overleftarrow{s}{}^{a}\right\vert _{\phi
,\pi ,\lambda }$ satisfying the algebra $\{\overleftarrow{U}{}^{a},%
\overleftarrow{U}{}^{b}\}=0$, and then in \cite{MR5} with arbitrary
parameters $\mu _{a}(\Gamma _{tr})$, including functionally-independent $\mu
_{a}(\phi ,\pi ,\lambda )$. The result is given by
\begin{align}
& \hspace{-0.7em}J_{\Lambda \overleftarrow{U}_{a}}\hspace{-0.1em}=\hspace{%
-0.1em}\mathrm{Sdet}\left\Vert \left[ {\Gamma _{tr}^{p}}{g}(\Lambda
\overleftarrow{U}_{a})\right] \overleftarrow{\partial }{}_{q}^{\Gamma
}\right\Vert \hspace{-0.1em}=\hspace{-0.1em}\exp \left[ -\left( \Delta
^{a}W\right) \mu _{a}-\textstyle\frac{1}{4}\left( \Delta ^{a}W\right)
\overleftarrow{s}_{a}\mu ^{2}\right] \big( 1-\textstyle\frac{1}{2}\Lambda
\overleftarrow{s}{}^{2}\big)^{-2}\hspace{-0.15em},  \label{superJaux6} \\
& \hspace{-0.7em}J_{\mu _{a}(\phi ,\pi ,\lambda )}=\exp \Big\{-\left( \Delta
^{a}W\right) \mu _{a}-\textstyle\frac{1}{4}\left( \Delta ^{a}W\right) {%
\overleftarrow{s}}_{a}\mu ^{2}-\mathrm{tr}\,\ln \left( e+{m}\right) \Big\},
\label{J-arb} \\
& \hspace{-0.7em}\ J_{\mu _{a}(\Gamma _{tr})}=\exp \Big\{-\mathrm{tr}\,\ln ( e+{m}) \Big\} g\big(\mu _{a}(\Gamma _{tr})\big)\exp\Big\{-( \Delta
^{a}W) \mu _{a}-\textstyle\frac{1}{4}( \Delta ^{a}W) {%
\overleftarrow{s}}_{a}\mu ^{2}\Big\}.  \label{J-arb-gen}
\end{align}%
The group-like element $g\big(\mu _{a}(\Gamma _{tr})\big)$  in (\ref{J-arb-gen}) draws
a difference between the Jacobians $J_{\mu _{a}(\phi ,\pi ,\lambda )}$ and
$J_{\mu _{a}(\Gamma _{tr})} $, because $\overleftarrow{s}_{a}$ are not reduced
to the nilpotent $\overleftarrow{U}_{a}$ as they act on $\Gamma _{tr}^{p}$.
In generalized Hamiltonian formalism, the Jacobians of corresponding
FD BRST-antiBRST transformations were calculated from first principles
by the rules (\ref{jacobianN2YM})--(\ref{jacobianN2YMgen}) in \cite{MR3,MR5}.

\section{On soft nilpotency and gauge-independent Standard Model with GZ horizon}
\label{Implication}

For FD parameters, finite BRST transformations allow one to obtain a new
form of the Ward identity and to establish the gauge-independence of the
path integral under a finite change of the gauge, $\Psi \rightarrow
\Psi+\Psi ^{\prime }$, provided that the SB BRST symmetry term $M=M_{\Psi }$
transforms to $M_{\Psi +\Psi ^{\prime }}=M_{\Psi }(1+\overleftarrow{s}\mu
(\Psi ^{\prime }))$, with $\mu (\Psi ^{\prime })$ being a solution of a
so-called compensation equation,
\begin{equation}
Z_{\Psi ,M_{\Psi }}(0,\phi ^{\ast })=Z_{\Psi +\Psi ^{\prime },M_{\Psi +\Psi
^{\prime }}}(0,\phi ^{\ast })\Rightarrow \Psi ^{\prime }(\phi ,\lambda |\mu
)=\textstyle\frac{\hbar }{i}\left[ \sum\nolimits_{n=1}\textstyle\frac{%
(-1)^{n-1}}{n}\left( \mu \overleftarrow{s}\right) {}^{n-1}\right] \mu,
\label{comeq}
\end{equation}%
for representations of the path integral (\ref{genfunbv}).
The Ward identity, depending on the FD parameter $\mu (\Psi ^{\prime })
=-\frac{i}{\hbar }g(y)\Psi ^{\prime }$, for $g(z)=1-\exp \{z\}/z$, $z\equiv ({%
i}/{\hbar })\Psi ^{\prime }\overleftarrow{s}$, and the gauge-dependence
problem are described by the respective expressions \cite{MR4}
\begin{equation}
\hspace{-0.7em}\textstyle\left\langle \hspace{-0.3em}\left\{ \hspace{-0.2em}1+\frac{i}{%
\hbar }\left[ \hspace{-0.1em}J_{A}\phi ^{A}+M_{\Psi }\hspace{-0.1em}\right]
\overleftarrow{s}\mu (\Psi ^{\prime })\hspace{-0.2em}\right\} \left( \hspace{%
-0.1em}1+\mu (\Psi ^{\prime })\overleftarrow{s}\hspace{-0.1em}\right) {}^{-1}%
\hspace{-0.3em}\right\rangle _{\Psi ,M,J}\hspace{-0.7em}=1\,\mathrm{and}%
\,\left\langle \hspace{-0.15em}(J_{A}\phi ^{A}+M_{\Psi })\overleftarrow{s}%
\hspace{-0.25em}\right\rangle _{\Psi ,M,J}\hspace{-0.3em}=0,  \label{mWIbvbr}
\end{equation}%
as one makes averaging with respect to $Z_{\Psi ,M_{\Psi }}(J,\phi ^{\ast })$.
The above equations are equivalent to those of \cite{Reshetnyak} in
the representation (\ref{genfun}) if we restrict ourselves by the set $G_w$
of $N=1$ BRST transformations with {soft nilpotency} imposed
on $\tilde{\mu}(\phi)$. Indeed, from (\ref{jacobweak}) and (2.40)
in \cite{Reshetnyak}, we find the compensation equation
\begin{eqnarray}
\imath \hbar\ln \big(1+\tilde{\mu}\overleftarrow{s}_{e} \big)
= \big(\exp\big\{ -[\Delta,\Psi^{\prime }]_{+}\big\}-1\big)S_{\Psi}, \label{compeqn}
\end{eqnarray}
whose resolvability implies that its right-hand side should be
\begin{eqnarray}
\overleftarrow{s}_{e}-\mathrm{closed }:   \big[\big(\exp\big\{ -[\Delta,\Psi^{\prime }]_{+} \big\}-1\big)S_\Psi\big]\overleftarrow{s}_{e} = 0. \label{rescompeqn}
\end{eqnarray}
For an infinitesimal change $\Psi^{\prime }$, this amounts to a \emph{soft nilpotency condition}:
$\Psi^{\prime}(\overleftarrow{s}_{e})^2 =0$.
For admissible changes of the gauge $\Psi^{\prime }$ satisfying this condition,
the solution to (\ref{compeqn}) for an unknown  $\tilde{\mu}$, with accuracy
up to a total derivative $(F\overleftarrow {\partial}_A)$, has the form
\begin{equation}\label{solncomp}
 \tilde{\mu} (\Psi ^{\prime }|\phi) = \frac{\Psi^{\prime }}{\Psi^{\prime }\overleftarrow{s}_{e}} \Big[\exp\Big\{\frac{\imath}{\hbar}\big(\exp\big\{ -[\Delta,\Psi^{\prime }]_{+} \big\}-1\big)S_\Psi \Big\}-1\Big].
\end{equation}
This allows one to obtain a new form of Ward identity depending on $\tilde{\mu} (\Psi ^{\prime })$
and to specify gauge dependence, with simiar results developed in \cite{Reshetnyak}.

$N=2$ FD BRST transformations solve the same problem under a finite
change of the gauge, $F\rightarrow F+F^{\prime }$, provided that the SB
BRST-antiBRST symmetry term $M_{F}$ transforms to $M_{F+F^{\prime }}
=M_{F}(1+\overleftarrow{s}^{a}\mu _{a}(F^{\prime })
+\frac{1}{4}\overleftarrow{s}^{2}\mu {}^{2}(F^{\prime }))$, with
$\mu _{a}(F^{\prime };\phi ,\pi ,\lambda)=\Lambda \overleftarrow{U}_{a}$
being a solution to the corresponding compensation equation based on (\ref{z(0)}):
\begin{equation}
Z_{F}(0)=Z_{F+F^{\prime }}(0)\Rightarrow F^{\prime }(\phi ,\pi ,\lambda |\mu
_{a})=4\imath {\hbar }\left[ \sum\nolimits_{n=1}\textstyle\frac{(-1)^{n-1}}{%
2^{n}n}\left( \Lambda \overleftarrow{U}^{2}\right) {}^{n-1}\Lambda \right] .
\label{comeqsp}
\end{equation}%
As a result, the Ward identity with the FD parameters $\mu
_{a}(F^{\prime })=\frac{i}{2\hbar }g(z)F^{\prime }\overleftarrow{U}_{a}$,
$\Lambda (\Gamma |{F}^{\prime })=\frac{i}{2\hbar }g(z){F}^{\prime }$ for $%
z\equiv ({i}/{4\hbar })F^{\prime }\overleftarrow{U}{}^{2}$,
acquires the form \cite{MR4}
\begin{eqnarray}
&&\textstyle\left\langle \left\{ 1+\frac{i}{\hbar }J_{A}\phi ^{A}\left[
\overleftarrow{U}^{a}\mu _{a}(\Lambda )+\frac{1}{4}\overleftarrow{U}^{2}\mu
^{2}(\Lambda )\right] -\frac{1}{4}\left( \frac{i}{\hbar }\right)
{}^{2}J_{A}\phi ^{A}\overleftarrow{U}^{a}J_{B}(\phi ^{B})\overleftarrow{U}%
_{a}\mu ^{2}(\Lambda )\right\} \right.  \label{mWI} \\
&&\quad \left. \times \textstyle\left( 1-\frac{1}{2}\Lambda \overleftarrow{U}%
^{2}\right) {}^{-2}\right\rangle_{F,J}=1, \nonumber
\end{eqnarray}%
and allows one to solve the gauge-dependence problem \cite{MR4} with a source-dependent
average expectation value with respect to $Z_{F}(J)$, corresponding to a gauge-fixing $F(\phi )$.

Using the $N=1, 2$ SB BRST symmetry term $M(\phi )$ as the horizon functional
$H(A)$ \cite{Zwanziger},
\begin{eqnarray}  \label{FuncM}
H(A)=\gamma^2\int d^Dx\big( d^Dy f^{mnk}A^n_{\mu}(x)(K^{-1})^{ml}(x,y)f^{ljk} A^{j\mu}(y)\ +\ D(\hat{N}^2{%
-}1)\big)
\end{eqnarray}
with the inverse Faddeev--Popov operator $(K^{-1})^{mn}(x,z)$ in Landau gauge, the Gribov mass $\gamma$,
and gauge-independent $Z_{H,\Psi }$, $Z_{H,F}$ in (\ref{comeq}), (\ref{comeqsp}), we find in a new gauge
\begin{equation}
\textstyle H_{\Psi ^{\prime }}(\phi )=H(A)\left\{ 1+\overleftarrow{s}\mu (\Psi ^{\prime
})\right\} \ \mathrm{or}\ H_{F^{\prime }}(\phi )=H(A)\left\{ 1+%
\overleftarrow{s}{}^{a}\mu _{a}(F^{\prime })+\frac{1}{4}\overleftarrow{s}{}%
^{2}\mu ^{2}(F^{\prime })\right\} .  \label{DeltaH}
\end{equation}%

At the same time, we can suggest a new $N=1$ or $N=2$ BRST-invariant and gauge-independent extension
of YM theory, by using a gauge-invariant horizon $H(A^h)=H(A)\big|_{A\to A^h}$ in terms
of gauge- and BRST-invariant transverse fields $A^h_{\mu{}}= (A^{h})^{n}_{\mu} T^n$, with $su(\hat{N})$
generators $T^n$ and a coupling constant $g$; see \cite{SemenovTyanshan}:
\begin{eqnarray}\label{gitrans}
&\hspace{-1.0em}&  \hspace{-1em} \textstyle A_{\mu}\hspace{-0.1em}= \hspace{-0.1em}A^h_{\mu}+  A^L_{\mu}:  A^h_{\mu}\hspace{-0.1em}
=\hspace{-0.1em}(\eta_{\mu\nu}- \frac{\partial_\mu\partial_\nu}{\partial^2})\Big(A^{\nu} -\imath g \big[\frac{\partial A}{\partial^2},
A^\nu - \frac{1}{2} \partial^\nu\frac{\partial A}{\partial^2}\big]  \Big)+ \mathcal{O}(A^3)\hspace{-0.1em} : \, A^h_{\mu}\overleftarrow{s}=0 ,\\
&\hspace{-1.0em}& \hspace{-1em} \textstyle H(A) = H(A^h)+ \gamma^2 \int d^Dx\, d^Dy\ R^m(A, \partial A, \partial_x;x,y)\partial^\mu A^m_\mu(y), \ \
H(A^h)\overleftarrow{s}=0,\label{gihor}
\end{eqnarray}
with a non-local function $R^m(x,y)$ in \cite{Pereira}. The structure of the second term in $H(A)$ allows one
to add it to the gauge term $B^m(\partial^\mu A^m_\mu) $ in the Faddeev--Popov action
$S_0 + \Psi \overleftarrow{s}$ (or the  $N=2$ BRST action $S_F$), in such a way that
the change of variables in $Z_{H,\Psi }$ is a  shift, $B^m \to B^m + \gamma^2 R^m$, with
the unity Jacobian completely eliminating the dependence on the SB BRST symmetry term in $Z_{H,\Psi }$.
Therefore, the action
\begin{equation}\label{brstinvgz} \textstyle  \hat{S}_{GZ}(\phi) \hspace{-0.1em}= \hspace{-0.1em}S_0
+ \int \hspace{-0.1em} d^Dx(\bar{C}{}^m \partial^\mu A^m_\mu) \overleftarrow{s} \hspace{-0.1em}+ H(A^h) ,\,
\mathrm{for } \, \phi^A\overleftarrow{s}\hspace{-0.1em}=\hspace{-0.1em}(D_{\mu }^{mn}C^{n}, -%
\textstyle\frac{1}{2}f^{mnl}C^{l}C^{n}, B^m,0)
\end{equation}
provides the gauge-independence in the YM theory and Standard Model \cite{MR5} under $R_\xi$-gauges,
with the same Faddeev--Popov operator $(K)^{mn}(x,y)$ and unaffected $N=1$
(with $\Psi \overleftarrow{s}$ replaced by $-\textstyle{\frac{1}{2}}F_\xi \overleftarrow{s}{}^2$, in (\ref{brstinvgz}) for $N=2$) BRST symmetry, for which one may expect the unitarity of the theory within
the Faddeev--Popov quantization rules \cite{FP}. The same results concerning the problems of unitarity and gauge-independence may be achieved within the local formulation of Gribov--Zwanziger theory \cite{Zwanziger}  when the horizon functional  is localized by means of a quartet of auxiliary fields $\phi_{\rm {aux}}=\big(\varphi^{mn}_\mu, \bar{\varphi}{}^{mn}_\mu;$  $\omega^{mn}_\mu, \bar{\omega}{}^{mn}_\mu \big)$,  having opposite Grasmann parities,  $\epsilon(\varphi, \bar{\varphi})= \epsilon(\omega, \bar{\omega})+1=0$, and being antisymmetric in $m,n$:
\begin{eqnarray}  \label{Sgamma}
  && \hat{S}_{\rm{GZ}}(\phi, \phi_{\rm {aux}})  \ = \  S_0(A)
+ \int \hspace{-0.1em} d^Dx(\bar{C}{}^m \partial^\mu A^m_\mu) \overleftarrow{s} \hspace{-0.1em}
+  S_{\gamma}(A^h,\phi_{\rm {aux}})\ , \\
 && S_{\gamma}\ = \ \int \hspace{-0.1em} d^Dx\Big({\bar \varphi}^{mn}_\mu K^{ml}(A^h) \varphi^{\mu{} l n}
- {\bar \omega}^{mn}_\mu K^{m l}(A^h) \omega^{\mu{} l n}  \label{Sgamma1} \\
&& \phantom{S_{\gamma}\ = \ } + \gamma\,f^{m nl}(A^h)^{\mu m}(\varphi_\mu^{nl}-\bar{\varphi}%
_\mu^{nl})+\gamma^2D
(\hat{N}{}^2-1)\Big)\,. \nonumber
\end{eqnarray}
The part $S_\gamma$ additional to the Faddeev--Popov action is explicitly $N=1$ BRST invariant, because of a trivial (vanishing) definition of $N=1$ ($N=2$) BRST transformations for the auxiliary fields: $\phi_{\rm {aux}} \overleftarrow{s}=0$ ($\phi_{\rm {aux}} \overleftarrow{s}{}^a=0$).\footnote{These results were obtained in June 2016, but were not included in the SFP'16 Proceedings due
to the limitations for a paper volume.}

Notice in conclusion that $N=1,2$ FD BRST transformations make it possible to study their explicit influence
on the Standard Model, reducibl€€e theoriess (such as the Freedman--Townsend model), the concept of average
effective action \cite{Reshetnyak,MR1,MR2,MR4,MR5}, and also allow one to extend themselves to the case
of  $N=m$ BRST transformations for arbitrary $m>2$, along the lines of \cite{UpReshMan}.

\paragraph{Acknowledgments}
The authors are grateful to the participants of the International Workshop
``Strong Field Problems in Quantum Field Theory'' (June 06--11, 2016, Tomsk, Russia)
for discussions. A.A.R. thanks A.D. Pereira for useful correspondence.
The study of P.Yu.M. is supported by the Tomsk State University Competitiveness
Improvement Program.

\end{document}